# Approaching Ferrite-Based Exchange-Coupled Nanocomposites as Permanent Magnets

Cecilia Granados-Miralles,[†] Matilde Saura-Múzquiz,[†] Henrik L. Andersen,[†] Adrián Quesada,[‡] Jakob V. Ahlburg,[†] Ann-Christin Dippel,[§] Emmanuel Canévet,[∥,⊥] and Mogens Christensen*[,†]

[†]Center for Materials Crystallography, Department of Chemistry and iNANO, Aarhus University, Langelandsgade 140, 8000 Aarhus, Denmark
[‡]Electroceramic Department, Instituto de Cerámica y Vidrio, CSIC, Kelsen 5, 28049 Madrid, Spain
[§]Deutsches Elektronen-Synchrotron (DESY), Photon Science, Notkestrasse 85, 22607 Hamburg, Germany
[∥]Laboratory for Neutron Scattering and Imaging, Paul Scherrer Institut (PSI), 5232 Villigen, Switzerland
[⊥]Department of Physics, Technical University of Denmark, 2800 Kgs. Lyngby, Denmark

*S* Supporting Information

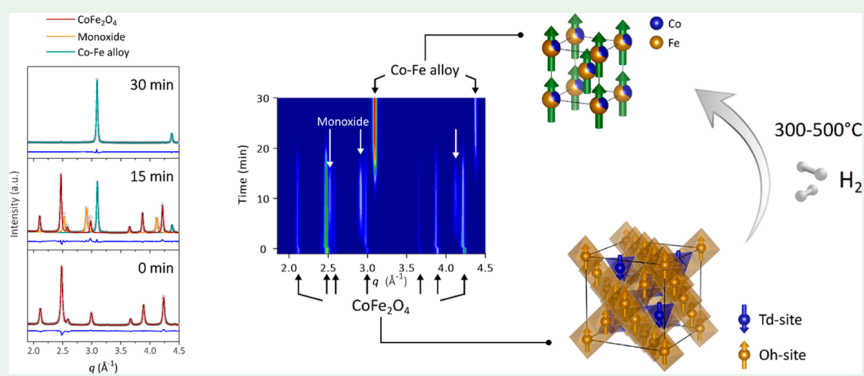

**ABSTRACT:** During the past decade, $CoFe_2O_4$ (hard)/Co−Fe alloy (soft) magnetic nanocomposites have been routinely prepared by partial reduction of $CoFe_2O_4$ nanoparticles. Monoxide (i.e., FeO or CoO) has often been detected as a byproduct of the reduction, although it remains unclear whether the formation of this phase occurs during the reduction itself or at a later stage. Here, a novel reaction cell was designed to monitor the reduction *in situ* using synchrotron powder X-ray diffraction (PXRD). Sequential Rietveld refinements of the *in situ* data yielded time-resolved information on the sample composition and confirmed that the monoxide is generated as an intermediate phase. The macroscopic magnetic properties of samples at different reduction stages were measured by means of vibrating sample magnetometry (VSM), revealing a magnetic softening with increasing soft phase content, which was too pronounced to be exclusively explained by the introduction of soft material in the system. The elemental compositions of the constituent phases were obtained from joint Rietveld refinements of *ex situ* high-resolution PXRD and neutron powder diffraction (NPD) data. It was found that the alloy has a tendency to emerge in a Co-rich form, inducing a Co deficiency on the remaining spinel phase, which can explain the early softening of the magnetic material.

## ■ INTRODUCTION

Permanent magnets (PMs) are present in countless applications, from everyday technology (computers, cell phones, speakers, microphones, household appliances, etc.) to industrial-scale energy-conversion and transportation devices (motors, generators, alternators, transformers, etc.).[1] They are also essential components in state-of-the-art technology dedicated to harvest renewable energy, e.g., wind, wave, or tidal power, as well as on electric vehicles.[2] Consequently, the optimization of PMs is not only necessary to keep up with the technological advances of our times but also a requirement on the road to sustainability, since the viability in the replacement of fossil fuels by green energy relies on our ability to fabricate lighter and more energy-efficient devices.[3]

The performance of a PM is usually evaluated based on its maximum energy product, $BH_{max}$. The $BH_{max}$ value depends on the magnetic field that the magnet is potentially able to produce (which is limited by the saturation magnetization, $M_s$) and on its resistance to demagnetization (i.e., coercivity, $H_c$). High $M_s$ and $H_c$ values are desirable for PMs in order to maximize their $BH_{max}$.

Unfortunately, these two properties do not usually coexist in single-phase materials.[4] For instance, the Co−Fe alloys are the materials with the highest potential magnetization known ($M_s$ = 240 A m$^2$/kg).[5] However, they have an effectively zero $BH_{max}$ as a consequence of the low $H_c$ values arising from their almost zero magnetocrystalline anisotropy. The combination of magnetic phases of different nature could potentially help breaking the natural constraints of single-phase materials.

Kneller and Hawig published the first model of a composite material combining a hard magnetic phase with large $H_c$ and a soft phase with a high $M_s$.[6] Their theoretical calculations predicted significant $BH_{max}$ enhancements in the composites with respect to the separate materials as long as the two magnetic phases were magnetically coupled at the atomic level, i.e., exchange-coupled. In broad strokes, they concluded that the requirements for improving the $BH_{max}$ through exchange-coupling are (i) an intimate contact between the two phases and (ii) a soft phase with crystallite sizes below a certain limit, usually on the order of a few tens of nanometers.[7,8]

The exchange-coupling theory laid the foundation of a completely new approach for producing high-performance PMs. Unfortunately, accomplishing an effective exchange-coupling between phases has proven more challenging in practice than in the theory. In the pursuit of a better understanding of the exchange-coupling phenomenon, a great amount of experimental and theoretical work has been dedicated to the subject during the past two decades, and it still remains an area of intensive research nowadays.[9−14] In particular, the $CoFe_2O_4$ (hard)/Co−Fe alloy (soft) composite has been assiduously studied over the past few years.[15−24] Besides cobalt−iron alloys having the largest $M_s$ known at room temperature,[5] the system has drawn special attention because it can be prepared by partial reduction of $CoFe_2O_4$. This chemical route directly leads to coexistence of the two magnetic phases, and consequently, a greater crystallographic coherency between them is expected compared with mixing independently synthesized species to make the composite.

The most extended strategy to prepare magnetic $CoFe_2O_4$/Co−Fe nanocomposites is a thermal treatment of $CoFe_2O_4$ nanoparticles in a $H_2$-rich atmosphere.[15−19] Other reduction agents have been used, e.g., activated charcoal[20] or $CaH_2$.[21] These composites have also been made in the shape of dense ceramic materials by means of spark plasma sintering (SPS).[23,24] Regardless of the preparation method, monoxides, i.e., FeO or CoO, have often been detected as impurities in $CoFe_2O_4$/Co−Fe composites prepared through partial reduction.[17,19,23,24] In none of the cases presented in the literature was it possible to determine whether the formation of this phase occurs during the reduction process (reaction intermediate) or at a later stage as a reoxidation triggered by the nanocomposites coming into contact with air. Soares et al. postulated a model to explain the temperature and field dependence of the nanocomposites and concluded that the presence of FeO has an influence on the magnetic properties.[25] An in-depth understanding of the preparation method is indispensable to gain control over the monoxide formation and, more importantly, to reach the optimal soft phase size and hard/soft composition. Here, we have addressed this matter by monitoring the reduction of $CoFe_2O_4$ nanoparticles using synchrotron radiation. *In situ* powder X-ray diffraction (PXRD) measurements during the reduction have yielded real time structural and microstructural information on the phases appearing and disappearing during the process.

Despite the large interest for the $CoFe_2O_4$/Co−Fe system, we find there is a general lack of quantitative analysis of the composition and its correlation to the magnetic properties of the sample. In the present work, quantitative information has been extracted from Rietveld refinements of neutron powder diffraction (NPD) and high-resolution PXRD data. Although NPD measurements are generally less accessible than PXRD, they are highly advantageous as they provide information on the magnetic structure of the materials, given that neutrons scatter from the atomic magnetic moments of the samples. Additionally, the neutron scattering lengths of Co and Fe are very different ($b_c$ = 9.45(2) and 2.49(2) fm, respectively),[26] ensuring good contrast between these two elements, unlike with X-rays. Magnetic hysteresis at room temperature has also been measured on the same samples using a vibrating sample magnetometer (VSM). The magnetic properties of several $CoFe_2O_4$/Co−Fe composites have been analyzed and discussed in the context of sample composition, crystallite size, and elemental composition of the individual phases.

## EXPERIMENTAL SECTION

**Synthesis of the Starting $CoFe_2O_4$ Material.** $CoFe_2O_4$ nanopowders with a volume-weighted average crystallite size of ≈14 nm were hydrothermally synthesized using the synthesis route described by Stingaciu et al.[27] A stoichiometric mixture of the metallic nitrates in aqueous solution was precipitated into a gel upon addition of a strongly alkaline solution (NaOH, 16 M) under constant magnetic stirring. Further details on the preparation of the precursor gel may be found in the Supporting Information. The as-prepared precursor was transferred to a 180 mL Teflon-lined stainless-steel autoclave, which was sealed and placed inside a Carbolite convection furnace preheated to 240 °C. After 2 h, the autoclave was removed from the furnace and left to cool in ambient conditions. The obtained suspension of nanoparticles was washed with ≈200 mL of deionized water and centrifuged at 2000 rpm for 3 min. The supernatant was discarded, and the remaining solid was washed with deionized water and centrifuged two more times. Finally, the product was dried in a vacuum oven (70 °C, 4 h), yielding about 9.3 g of nanosized $CoFe_2O_4$ (reaction yield ≈96%). Relatively narrow crystallite size distributions are expected based on previous investigations by Andersen et al. on the hydrothermal synthesis of $CoFe_2O_4$ under similar conditions.[28]

***In Situ* PXRD Studies.** *Reduction of the $CoFe_2O_4$ Nanoparticles.* Reduction experiments were carried out using a custom-made reduction cell optimized for *in situ* PXRD experiments. Figure 1 shows an illustration of the *in situ* reduction setup. A small amount of $CoFe_2O_4$ nanopowders (≤ 10 mg) was loaded into a 45−50 mm long fused-silica capillary with both ends open (i.d. = 0.70 mm, o.d. = 0.85 mm). A piece of heat-resistant polyamide tubing was introduced through each end of the fused-silica capillary, applying a gentle pressure to confine the powders in the middle region of the capillary (≈10 mm). The polyamide tube was mechanically twisted and turned beforehand to help it act as traps to prevent the powders from escaping the capillary when flowing gas through the system. The loaded capillary was sealed using Swagelok fittings, as shown in Figure 1, and a controlled flow (5–30 mL/min) of a reducing gas mixture 4% $H_2$/Ar was run through the system. Subsequently, the sample was subjected to elevated temperatures (300–500 °C). A hot-air stream (20 L/min) generated by a commercial heating gun (Hi-Heater 440 W, $\phi$ = 13 mm, Miyakawa Corporation) was directed toward the capillary. A 20 mm wide quartz nozzle was attached at the top end of the blower to ensure a homogeneous heating of the entire sample. Very fast heating rates were achieved by this method, with the set temperatures reached within the first 15 s of heating in all cases.

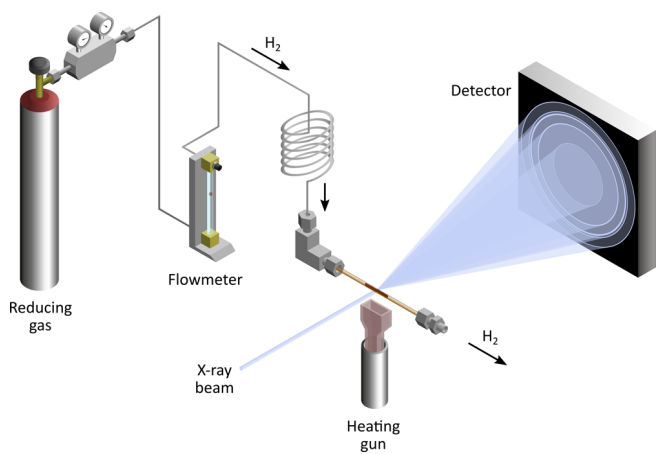

**Figure 1.** Illustration of the *in situ* reduction setup.

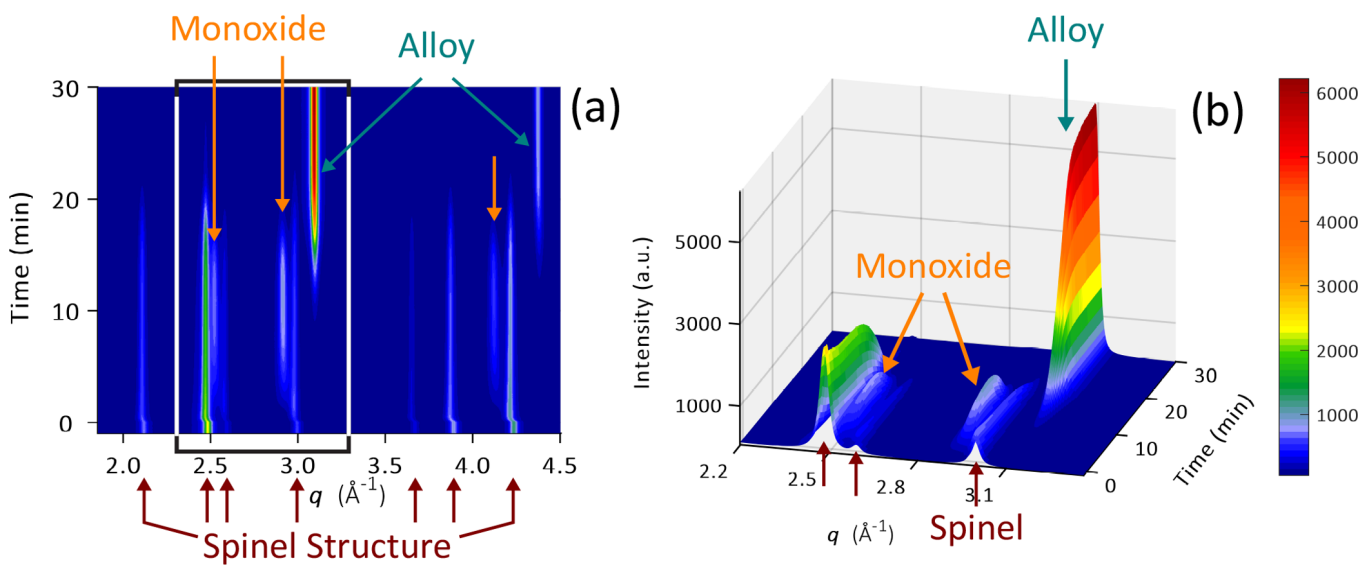

**Figure 2.** (a) Contour plot of the time-resolved PXRD data set from the reduction experiment performed at 400 °C and 10 mL/min of 4% $H_2$/Ar. A constant gas flow was maintained during the entire data collection, while heating was started at time = 0. For clarity, only the $q$-region 1.85–4.5 Å$^{-1}$ is shown here, although data were collected up to 9.5 Å$^{-1}$. (b) Selected $q$-region of the same data set plotted using a 3D-view. The brown, orange, and turquoise arrows indicate the spinel, monoxide, and alloy reflections, respectively.

*In Situ Powder Diffraction Measurements.* The aforementioned reduction cell was mounted at the high-resolution powder diffraction beamline P02.1 at the PETRA III synchrotron (DESY, Hamburg).[29] Reduction experiments were carried out while being monitored using synchrotron radiation with a wavelength of 0.207 00 Å and a beam size of 0.5 × 0.5 mm$^2$. PXRD data with a time resolution of 5 s were measured up to $2\theta$ = 18° (i.e., $q$ = 9.5 Å$^{-1}$ at the given wavelength) until the reduction was complete.

The time-resolved diffraction data were recorded using a fast, amorphous silicon area-detector PerkinElmer XRD1621 (2048 × 2048 pixels, pixel size 200 × 200 μm$^2$) located ca. 915 mm behind the sample. The collected 2D-images were azimuthally integrated to 1D-patterns using the software Dioptas.[30] The sample-to-detector distance, beam-center position, and detector tilt were extracted using the PXRD data collected for a standard powder (NIST LaB$_6$ SRM 660b)[31] packed in a 0.7 mm quartz capillary. The data on the standard powder were measured in the same experimental configuration as the samples. Details about the data integration and representative examples of the 2D-data collected *in situ* may be found in the Supporting Information. For an extended description of the *in situ* PXRD data treatment procedure employed, the reader is referred to the article by Andersen et al.[32]

**Reduction at a Larger Scale and *ex Situ* Characterization.** *Preparation of the Nanocomposites: Partial Reduction.* About 2 g of $CoFe_2O_4$ nanoparticles were spread on an $Al_2O_3$ crucible with approximate dimensions 6 × 4 cm$^2$, which was placed at the hottest spot of a tubular furnace (C.H.E.S.A. Ovens). After the furnace was sealed at both ends, it was purged and evacuated to a pressure of approximately 10$^{-2}$ mbar using a vacuum pump connected at the furnace outlet. A gas mixture 10% $H_2$/$N_2$ was regulated to flow through the furnace and produce a gas pressure of 20 mbar inside the furnace. Once the pressure was stable, the thermal treatment was initiated. An initial heating ramp of 100 °C/min was used to drive the temperature up to the set point (350–450 °C), the temperature at which the system was maintained for 2 h. Afterward, the sample was left to cool inside the furnace, maintaining the flow of reducing gas. Once the temperature was below 75 °C, air was let inside the system and the sample was removed from the furnace.

*Ex Situ X-Ray and Neutron Powder Diffraction Measurements.* PXRD data were collected using a Rigaku SmartLab diffractometer in Bragg–Brentano $\theta/\theta$ geometry (incident-slit opening = of 1/2°) with a diffracted beam monochromator (DBM) in front of a D/teX Ultra detector. For each sample, two independent data sets were measured using X-rays generated by different anodes, i.e., Cu K$\alpha$ ($\lambda_{Cu\ K\alpha 1}$ = 1.540 593 Å, $\lambda_{Cu\ K\alpha 2}$ = 1.544 427 Å) and Co K$\alpha$ ($\lambda_{Co\ K\alpha 1}$ = 1.789 00 Å, $\lambda_{Co\ K\alpha 2}$ = 1.792 84 Å). Data were collected in the $q$-range 1.0–6.6 Å$^{-1}$ at both wavelengths, and the instrument was operated at 40 kV and 180 mA and at 35 kV and 170 mA, respectively.

NPD data were collected for all the samples at the Cold Neutron Powder Diffractometer, DMC,[33] at the Swiss Spallation Neutron Source, SINQ (Paul Scherrer Institut, PSI, Villigen, Switzerland) using a wavelength of 2.458 97(11) Å and in the $q$-range 0.5–3.7 Å$^{-1}$. NPD data over a wider $q$-range (0.3–8.3 Å$^{-1}$) were additionally collected for the partially reduced composites, at the High Resolution Powder diffractometer for Thermal neutrons, HRPT,[34] at SINQ, using a wavelength of 1.493 65(7) Å.

*Vibrating Sample Magnetometry.* A small fraction of each sample (mass = 10–15 mg) was gently compacted into a cylindrical pellet (diameter = 3.00 mm, thickness = 0.60–0.70 mm) using a hand-held press. The pellet mass was determined with a precision of 0.001 mg after being dried in a vacuum furnace (1 h, 60 °C). Field-dependent magnetization curves were measured at 300 K as a function of an externally applied field, $H_{app}$, using a vibrating sample magnetometer (VSM option for the Physical Property Measurement System PPMS, Quantum Design). $H_{app}$ was applied along the direction normal to the pellet surface and in the range ±2 T (≈ ±1590 kA/m).

## ■ RESULTS AND DISCUSSION

**Sequential Rietveld Refinements of *in Situ* Synchrotron PXRD.** Seven *in situ* reduction experiments were carried out using synchrotron PXRD to evaluate the influence of the gas flow (5, 10, 20, and 30 mL/min at 400 °C) and the temperature (300, 350, 400, and 500 °C at 10 mL/min). $CoFe_2O_4$ from the same synthesis batch was used as starting material for all experiments.

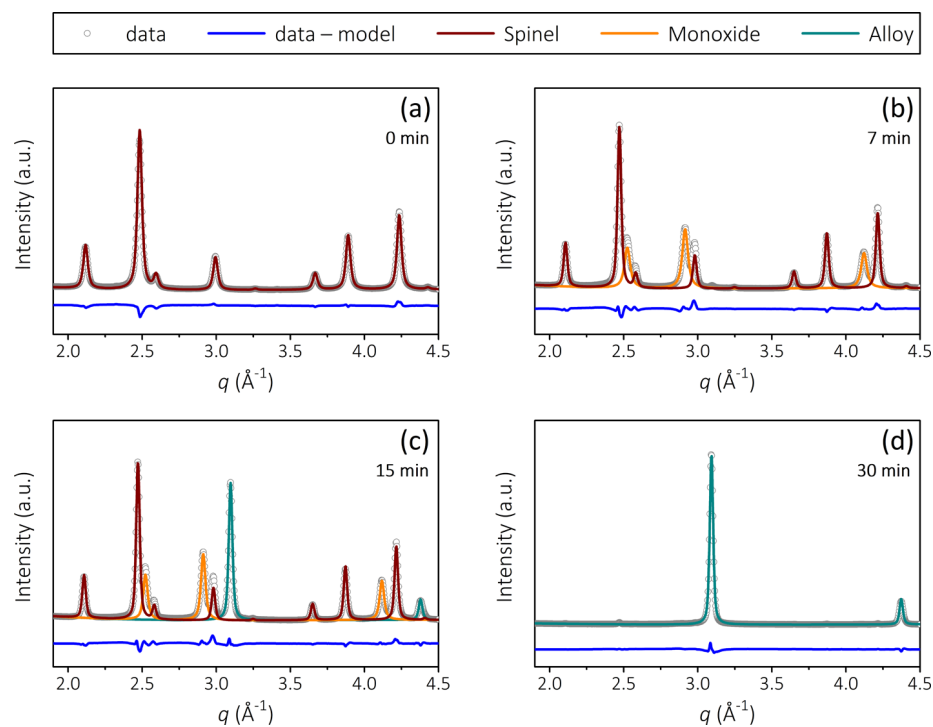

**Figure 3.** PXRD data collected at (a) time = 0, (b) 7 min, (c) 15 min, and (d) 30 min during the reduction experiment carried out at 400 °C and 10 mL/min, along with the corresponding model for each of the phases present. The open gray circles show the experimental data. The superimposed lines represent the refined Rietveld models for the spinel (brown), the monoxide (orange), and the alloy (turquoise), modeled as $CoFe_2O_4$, $Co_{0.33}Fe_{0.67}O$, and $Co_{0.67}Fe_{1.33}$, respectively. The blue line at the bottom of each graph is the difference between the experimental data and total Rietveld model.

A representative example of the diffraction data collected during these *in situ* experiments is shown in Figure 2a. At the beginning of the experiment, reducing gas is running through the sample at room temperature. The corresponding diffraction data (negative times) exhibit the characteristic pattern of a pure spinel structure, such as $CoFe_2O_4$. As soon as heating starts (time = 0), an abrupt shift of the pattern toward lower $q$-values is observed. This shift reflects the unit cell expansion upon heating. After 30 min of reducing treatment at these conditions (i.e., 10 mL/min, 400 °C), the initial oxide is fully reduced into a metallic alloy with a body-centered cubic (bcc) crystal structure.

The time-resolved diffraction data collected during reduction reveal that the transformation from the spinel to the alloy does not take place directly, but through an intermediate phase, which is indexed as a metal(II) oxide or monoxide. The formation of monoxide as an intermediate was observed for all the reduction experiments conducted, albeit at different speeds depending on the specific conditions of gas flow and temperature. The data inside the rectangular area in Figure 2a are represented in Figure 2b using a 3D-perspective, which shows more clearly the gradual appearance and disappearance of the monoxide phases.

Rietveld analysis of the diffraction data collected *in situ* was carried out using the software FullProf,[35] assuming a 1:2 Co:Fe stoichiometry for all phases. Thus, the spinel was modeled as $CoFe_2O_4$ ($Fd$-$3m$), the monoxide as $Co_{0.33}Fe_{0.67}O$ ($Fm$-$3m$), and the alloy as $Co_{0.67}Fe_{1.33}$ ($Pm$-$3m$). The site occupancies of the atoms were not refined given that Co and Fe are practically indistinguishable by X-ray diffraction (see Supporting Information).

Figure 3 shows the Rietveld models refined for four different frames selected from the diffraction data displayed in Figure 2. Single-phase spinel is found before the reduction starts (time = 0). After 7 min of heating, the spinel coexists with a monoxide phase, while a hint of the alloy is already observable at $q \approx 3.1$ Å$^{-1}$. The three phases are present simultaneously at intermediate times (15 min). At the end of the experiment (30 min), the two oxides have practically disappeared (spinel ≤2.0(3) wt %, monoxide ≤1.0(3) wt %), while the metallic alloy accounts for the 97.1(5) wt % of the sample.

Rietveld refinements were run sequentially on the time-resolved diffraction data sets, yielding refined values for the weight fractions, unit cell parameters, and crystallite sizes of the different phases as a function of time. Further information on these refinements is given in the Supporting Information.

*Influence of the Reducing Gas Flow.* Plotted in Figure 4 are the refined parameters corresponding to four different reduction experiments carried out at 400 °C and variable gas flows, i.e., 5, 10, 20, and 30 mL/min. The obtained results show that the gas flow has a clear influence on the phase composition (see Figure 4a–c). Thus, the alloy first appeared after about 15 min using a flow of 5 mL/min, while it took less than 5 min to form with a flow of 30 mL/min. The time required for full conversion from spinel to alloy ranged from 10 to 40 min depending on whether the highest or the lowest flow was used, respectively. Regardless of the flow, the monoxide formed almost instantaneously, although its lifetime varied between 10 and 30 min from the lowest to the highest flow. In all cases, the monoxide formed and vanished during the experiment, which confirms its role as an intermediate in the reduction process. The same chemical process was observed in all cases, but taking place at a faster speed for higher gas flow rates. It is therefore concluded that the $H_2$ availability in the system is a limiting factor for the reduction kinetics at the given temperature.

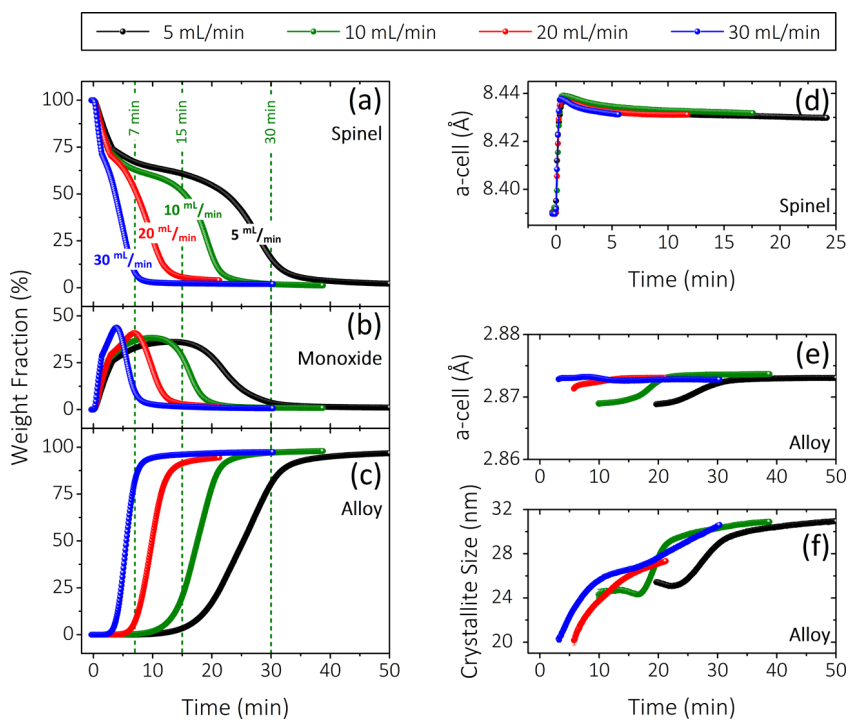

**Figure 4.** Results obtained from sequential Rietveld refinements of the *in situ* PXRD data collected during the reduction experiments conducted at 400 °C and variable gas flows: 5 mL/min (black), 10 mL/min (green), 20 mL/min (red), and 30 mL/min (blue). (a–c) Weight fractions for the three refined phases. (d) Unit cell parameter for the spinel. (e) Unit cell parameter and (f) volume-averaged crystallite size of the metallic alloy.

Figure 4d shows the unit cell parameter of the spinel phase. The fast unit cell expansion observed initially is followed by a contraction that gradually continues down to a final value of ≈8.43 nm, regardless of the gas flow. Figure 4e shows the unit cell of the metallic alloy as a function of time. In all cases, the unit cell increases with increasing times until it reaches a final value of ≈2.873 nm for all four experiments. This time-dependent increase on the alloy cell parameter could simply be an effect of the phase growth dynamics, but it could also be reflecting variations over time in the elemental composition of the alloy, i.e., the Co:Fe atomic ratio. The experiments described in the succeeding sections shed more light on this observation.

The crystallite sizes of both oxides (not shown) increase steadily until the phases start to disappear. On the other hand, the crystallite growth of the alloy shows a pronounced discontinuity that seems to coincide in time with the completion of the reduction (see Figure 4f). The two-step character of the growth curve suggests different growth-limiting mechanisms in each of the steps. During the first step, oxide crystallites turn purely metallic; i.e., the alloy growth must be due to the reduction process advancing. Only the alloy is present during the second step, and the alloy growth in this case is attributed to an Oswald ripening of the crystallites, induced by the elevated temperature.

*Influence of the Temperature.* To evaluate the influence of the temperature on the reduction process, three additional experiments were carried out at 300, 350, and 500 °C, while keeping the gas flow fixed at 10 mL/min. The refined parameters for the three corresponding diffraction data sets are plotted in Figure 5, along with those corresponding to the experiment at 400 °C and 10 mL/min (also represented in Figure 4 in green color).

The refined weight fractions (see Figure 5a–c) reveal that a higher temperature causes a faster reduction. The process is considerably slower at the lowest temperature, 300 °C, which is plotted in gray and using a 5 times longer time scale (top *x*-axis). As observed on the gas flow series of experiments, the monoxide is found as an intermediate at all studied temperatures. This monoxide phase is seen to disappear at shorter times than the spinel. This is especially visible at the lowest temperatures, but the same seems to occur at 400 and 500 °C. Therefore, it is possible to obtain $CoFe_2O_4$/Co–Fe composites free of monoxide but only in a limited range of reaction times, this time interval being shorter the higher the temperature, as the whole process is speeded up. Thus, in terms of designing an optimized synthesis route, it should be noted that lower reduction temperatures are more likely to yield monoxide-free composites.

Figure 5d–f shows the volume-averaged crystallite sizes for the temperature series. A faster crystallite growth is expected at elevated temperatures, and this is indeed observed for all three phases. When monitoring dynamic processes in which different phases coexist and evolve, a decrease in size is often observed coinciding with phase extinctions, as the crystallites of the specific phase are consumed.[36] This is especially visible here for the monoxide (see Figure 5e). The two-step growth of the alloy crystallites seen in the gas flow study also takes place in the temperature series.

The refined unit cell parameters are plotted as a function of time in Figure 5g–i. For the three phases, the trends observed here are similar to those seen in the gas flow series. However, the absolute values are temperature-dependent: the higher the temperature, the larger the unit cell, which is attributed to thermal expansion. The alloy unit cell ceases its expansion as soon as it becomes the sole phase present in the system. The initial increase can be explained by changes in the elemental composition of the phase. The lattice parameter of bcc-based Co–Fe alloys increases with increasing Fe content.[37] Therefore, the unit cell expansions would be explained by an increasing Fe content as the reduction progresses. Once the sample is fully reduced, the alloy composition remains stable at the Co:Fe ratio of the original material (ideally 1:2). Similar compositional changes would explain the unit cell trends refined for the monoxide. Although the range of cell parameter values found in the literature for the FeO and CoO is relatively broad depending on size and stoichiometry ($a$(FeO) = 4.280−4.326 Å[38−40] and $a$(CoO) = 4.240−4.273 Å,[41−43] at room temperature), the values for FeO are always larger than for CoO. Consequently, the increasing unit cell suggests that a Co-rich monoxide is obtained initially, followed by increasing Fe incorporation.

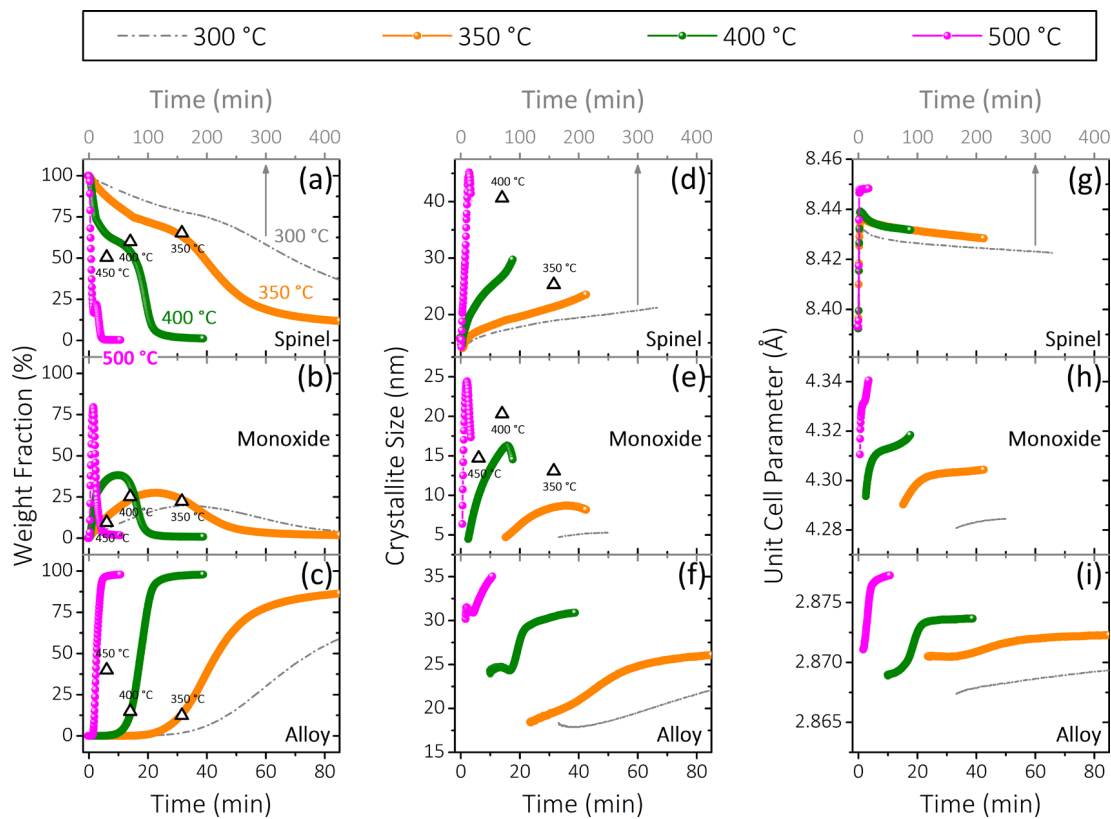

**Figure 5.** Data plotted in color (*in situ*): (a−c) Weight fractions, (d−f) volume-averaged crystallite sizes, and (g−i) unit cell parameters obtained from sequential Rietveld refinements of the *in situ* PXRD data collected during reduction using a gas flow of 10 mL/min and variable temperatures: 300 °C (gray), 350 °C (orange), 400 °C (green), and 500 °C (pink). The 300 °C experiment is plotted on a 5 times longer time scale (top *x*-axis). The 400 °C experiment (green) is the same as the one shown in the same color in Figure 4. Black open triangles (*ex situ*): (a−c) Weight fractions and (d−f) crystallite sizes corresponding to nanocomposites prepared *ex situ* at 350, 400, and 450 °C (plotted at time = 31.5, 14, and 6 min, respectively). Uncertainties smaller than symbol size. The missing values are outside of the range plotted in the graph. The reader is referred to the next section for further details on the *ex situ* experiments.

The formation of Co-rich reduced phases (i.e., monoxide and alloy) necessarily implies a Co deficiency on the remaining unreduced spinel. In other words, the starting cobalt spinel, $CoFe_2O_4$, is partially turned into an iron spinel oxide, e.g., $\gamma$-$Fe_2O_3$ (maghemite) or $Fe_3O_4$ (magnetite). After the initial heating-motivated cell expansion, a moderate decrease in cell parameter is registered for the spinel. This would in principle tip the scales in favor of $\gamma$-$Fe_2O_3$, considering that $a(\gamma$-$Fe_2O_3)$ = 8.34 Å < $a(CoFe_2O_4)$ = 8.39 Å < $a(Fe_3O_4)$ = 8.40 Å (values for bulk phases at room temperature).[5] However, the differences are too small to draw any conclusions based on the cell parameter alone, as the cell dimensions in nanoparticles may change due to finite size and strain effects. As may be seen from Figure 5h,i, the cell parameters of the reduced phases spread over a wider range the higher the temperature is. This suggests that at low temperatures the reduced phases arise closer to stoichiometry, and the Co deficiency on the spinel is presumably less pronounced.

Hence, the stoichiometry of the constituent phases can be controlled by tuning the experimental parameters. Based on our *in situ* investigations, a specific composition is achieved faster by increasing the temperature. However, the higher the temperature, the more the stoichiometry of the reduced phases will deviate from that of the starting material. That particular composition can also be obtained at lower temperatures, at the cost of increasing the treatment duration, and in this case the change in stoichiometry is minimized.

**Joint Rietveld Refinements of *ex Situ* PXRD and NPD.** Three nanocomposites, of different compositions and crystallite sizes, were prepared by partial reduction of $CoFe_2O_4$ nanoparticles in three independent reduction treatments at 350, 400, and 450 °C, respectively. PXRD and NPD data were collected on these composites and on the starting $CoFe_2O_4$ material, and a Rietveld model was built for each sample. The model was refined simultaneously against all the independent powder diffraction patterns collected for each sample, i.e., four patterns in the case of the composites and three for the starting material.

Figure 6 shows NPD patterns collected using two different instruments for the nanocomposite prepared at 350 °C, along with the corresponding Rietveld models. The total model is represented in black, while the red line corresponds to the magnetic contribution alone. In order to build robust and physically plausible

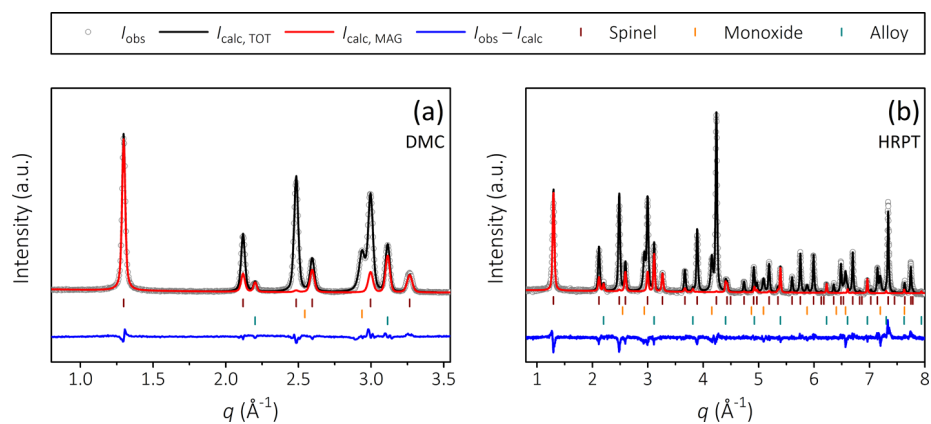

**Figure 6.** NPD data collected at (a) DMC and (b) HRPT for the nanocomposite prepared at 350 °C in the tubular furnace along with the corresponding Rietveld models. The open gray circles show the experimental data, the black line represents the total model, and the red line corresponds to the magnetic contribution alone. The Bragg positions of the different phases present are represented by the vertical ticks underneath the patterns, in brown color for the spinel, orange for the monoxide, and turquoise for the alloy.

Table 1. Results from the Joint Rietveld Refinements of the *ex Situ* and Neutron Powder Diffraction Data

| | spinel | | | monoxide | | | alloy | | |
|---|---|---|---|---|---|---|---|---|---|
| sample | weight fraction (%) | crystallite size (nm) | refined elemental composition | weight fraction (%) | crystallite size (nm) | refined elemental composition | weight fraction (%) | crystallite size (nm) | refined elemental composition |
| starting material | 100.0(1) | 13.34(3) | $Co_{0.90(2)}Fe_{2.10(2)}O_4$ | | | | | | |
| 350 °C | 65.3(2) | 25.3(1) | $Co_{0.56(3)}Fe_{2.44(3)}O_4$ | 22.36(7) | 13.1(1) | $Co_{0.53(1)}Fe_{0.47(1)}O$ | 12.39(4) | 46.1(5) | $Co_{0.88(2)}Fe_{1.12(2)}$ |
| 400 °C | 60.0(1) | 40.6(2) | $Co_{0.44(3)}Fe_{2.56(3)}O_4$ | 25.23(5) | 20.3(2) | $Co_{0.49(1)}Fe_{0.51(1)}O$ | 14.80(3) | 70.9(8) | $Co_{1.04(2)}Fe_{0.96(2)}$ |
| 450 °C | 50.5(1) | 81.4(1) | $Co_{0.23(7)}Fe_{2.77(7)}O_4$ | 9.50(3) | 14.7(4) | $Co_{0.12(3)}Fe_{0.88(3)}O$ | 40.03(9) | 50.0(4) | $Co_{1.10(2)}Fe_{0.90(2)}$ |

*The uncertainties shown in parentheses in the table are calculated based on the propagation of the uncertainties of the refined parameters, and they represent the minimum uncertainty the calculated values may have.

Rietveld models for the samples, a number of constraints were introduced in the joint refinements of the models. A detailed description of the Rietveld analysis of these data may be found in the Supporting Information. Table 1 summarizes the parameters of interest obtained from the joint Rietveld refinements.

*Sample Composition: Weight Fractions.* The starting material used to prepare the nanocomposites was phase-pure spinel. The produced nanocomposites had different reduction degrees—the reduction degree being defined as the fraction of metallic alloy present in the sample (i.e., alloy wt %). The alloy only represented 12.39(4) wt % and 14.80(3) wt % in the nanocomposites prepared at 350 and 400 °C, respectively, while the reduction treatment at 450 °C led to a much higher reduction degree (alloy wt % = 40.03(9)). Considering that these *ex situ* treatments had a duration of 2 h, the achieved reduction degrees were significantly lower than expected based on the shorter *in situ* experiments.

The sample obtained from the *ex situ* treatment at 350 °C had a comparable composition to the *in situ* sample reacted at the same temperature after only ≈31.5 min of experiment. This timestamp was graphically estimated by plotting the weight fractions refined for the *ex situ* sample on top of the values refined for the *in situ* experiment at the same temperature (see the open triangles tagged as "350 °C" in Figure 5a–c). With respect to the 400 °C *ex situ* sample, a comparable reduction degree was obtained *in situ* at this temperature after only ≈14 min. Although there is no equivalent *in situ* experiment to the *ex situ* one performed at 450 °C, the corresponding data were plotted at time = 6 min. The timestamps estimated to plot the *ex situ* results together with the *in situ* parameters are only meant for qualitative comparison of the two experimental setups. The offset in time between *ex situ* and *in situ* experiments comes from the differences between the experimental setups. (i) The availability of reducing gas might be a limiting factor *ex situ*, given that the amount of starting material is approximately 200 times larger. (ii) The likelihood of the reducing gas reaching the powders is substantially higher *in situ*, since the gas flows directly through the capillary. (iii) The heating of the sample in the furnace (*ex situ*) is significantly slower than the direct heating provided by the heat gun (*in situ*). (iv) Additionally, there might be a temperature offset derived from the way the temperature is measured in each case: directly on the capillary (*in situ*) and on the outer surface of the quartz tube from the furnace (*ex situ*). As shown in the *in situ* experiments, low gas availability and low temperatures cause a slowdown of the reduction process. Therefore, due to the aforementioned differences between the two setups, a substantial decrease in reduction speed is expected *ex situ* compared to *in situ* experiments.

Previous studies have shown that it is possible to avoid the presence of monoxide in the final product.[19] In the cited study, monoxide-free composites were obtained using the same furnace, after only 30 min at 400 °C and the same gas pressure used here (20 mbar), but using 10 times less sample (0.2 g). However, for the 2 g of sample prepared here (required to perform NPD measurements), 2 h at 450 °C was still not enough to avoid the presence of the monoxide.

*Crystallite Size of the Constituent Phases.* The refined volume-weighted average crystallite size obtained for the starting material was 13.34(3) nm. The crystallite sizes refined for each of the three nanocomposites differ substantially depending on the preparation temperature. The refined values are plotted in Figure 5g−i using the same timestamps derived from the weight fractions in Figure 5a−c. For all temperatures and phases, the sizes refined for the *ex situ* samples are considerably larger than the corresponding *in situ* ones as a consequence of the prolonged heating times required when using the *ex situ* setup.

The spinel phase grows in crystallite size as the preparation temperature increases. The same would be expected for the alloy, but the sample prepared at the highest temperature falls outside of this trend. However, the smaller size refined for 450 °C could be an artifact originating from the description of the phase in the Rietveld model. Previous studies on this system have shown that the alloy tends to segregate in two or more distinct phases, this effect becoming more pronounced at higher temperatures.[17,19] The different alloy phases have the same parent structures but slightly different unit cell dimensions, which produces diffraction patterns with severe peak overlap. Despite the previously mentioned observations, the alloy was described as a single phase here. This approximation holds well for the low temperatures, but it leads to a deficient description of the peaks width for the 450 °C sample. For the latter, the model tends toward artificially broadened profiles aiming to describe the overlapping peaks in the data. Although approximating the alloy to a single phase causes an underestimation of the size for the highest temperature composite, it is a necessary compromise for a meaningful Rietveld analysis of these data.

*Elemental Composition of the Constituent Phases.* The results from joint Rietveld analysis unambiguously show that the elemental composition of the different phases changes as a function of the reduction degree. The distribution of the metallic cations/atoms among the different phases is represented in Figure 7 for all samples. The elemental composition of the starting material, i.e., $Co_{0.90(2)}Fe_{2.10(2)}O_4$, differed slightly from the Co:Fe ratio of 1:2 expected from a stoichiometric Co spinel. Consequently, the refined atomic fraction of Co with respect to the total metal content in the starting material was 30.1(8) at. % instead of the expected 33.3 at. %. The Co content of the spinel phase in the three nanocomposites (350 °C, 400 °C, and 450 °C) is smaller than that of the starting material, and it is further diminished as the reduction advances (i.e., with increasing temperature). It is therefore concluded that the Co is preferentially removed from the spinel structure during the reduction, leading to reduced species rich in Co. This result is in agreement with the *in situ* observations.

For the nanocomposite prepared at 350 °C, the monoxide shows a Co surplus with respect to the stoichiometry of the starting material. This indicates that Co is more prone to form the monoxide than Fe. For the alloy, the Co content is very similar to that in the monoxide (approximately 50%). Therefore, from observation of this sample alone, it is not clear whether there is any preference between the two elements when it comes to the alloy formation. For the 400 °C sample, the distribution of the metallic species among the phases is very similar to 350 °C. However, a significant difference is observed after treatment at 450 °C: the monoxide becomes Co-deficient compared to the initial stoichiometry, while the alloy remains significantly Co-rich (see horizontal white line in Figure 7). This suggests that $Co^{2+}$ is more easily reduced than $Fe^{2+}$, which is congruent with the reduction potentials tabulated for these cations ($E°(Co^{2+}) = -0.28$ eV > $E°(Fe^{2+}) = -0.447$ eV)[44]—although these are only meant for reference, as they are exclusively valid for the cations in solution.

The propensity of the monoxide to be rich in Co is also reasonable from the electrochemical point of view, as the $Co^{2+}$ in the spinel does not need to change oxidation state to form CoO, while the $Fe^{3+}$ needs to be reduced to $Fe^{2+}$ first. In the case of the spinel, two options are contemplated. If $CoFe_2O_4$ turns into $\gamma\text{-}Fe_2O_3$, two-thirds of the $Co^{2+}$ would be replaced by $Fe^{3+}$, while the remaining one-third would stay vacant to preserve charge neutrality.[45] Therefore, this transformation would not involve reduction of any species (see eq 1). On the other hand, for $CoFe_2O_4$ to become $Fe_3O_4$, some of the $Fe^{3+}$ has to be reduced to $Fe^{2+}$ for replacing $Co^{2+}$ in the structure (see eq 2). The superscripted roman numbers in eqs 1 and 2 indicate the oxidation states of the metallic atoms in the different compounds.

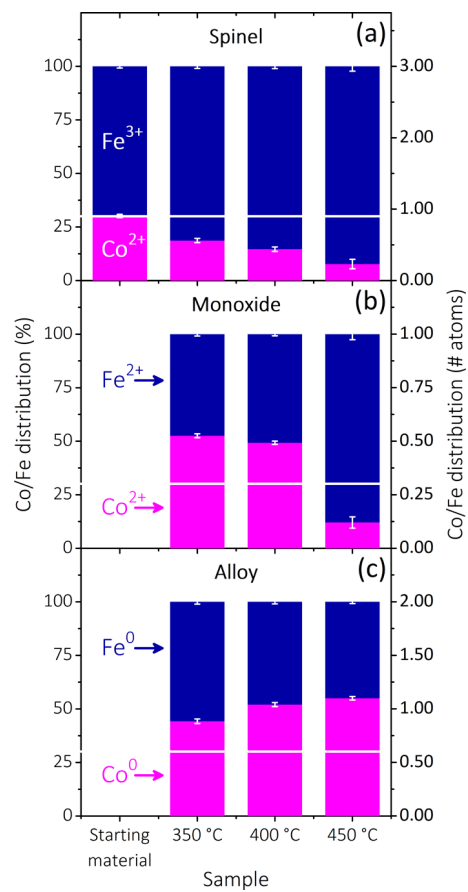

Figure 7. Distribution of the Co (pink) and Fe (dark blue) cations/atoms among the crystallographic sites available for metallic elements in the (a) spinel, (b) monoxide, and (c) alloy structures. The horizontal white line indicates the random distribution of Co and Fe for the stoichiometry refined for the starting material, i.e., 0.90(2):2.10(2).

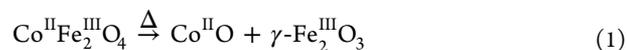

$$Co^{II}Fe_2^{III}O_4 \xrightarrow{\Delta} Co^{II}O + \gamma\text{-}Fe_2^{III}O_3 \qquad (1)$$

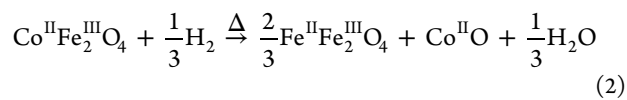

$$Co^{II}Fe_2^{III}O_4 + \frac{1}{3}H_2 \xrightarrow{\Delta} \frac{2}{3}Fe^{II}Fe_2^{III}O_4 + Co^{II}O + \frac{1}{3}H_2O \qquad (2)$$

Unfortunately, none of the experiments performed in this work have been able to verify whether the formation of $Fe_3O_4$ during reduction is more likely than $\gamma\text{-}Fe_2O_3$ or vice versa. A thermal treatment of the same starting material in a nonreducing atmosphere (2 h, 350 °C, pure $N_2$) did not produce any monoxide; i.e., the process predicted by eq 1 did not take place spontaneously. However, this does not disprove the formation of $\gamma\text{-}Fe_2O_3$ in reducing conditions.

*Magnetic Properties at Room Temperature.* The magnetic hysteresis was measured for the starting material and for the three nanocomposites prepared *ex situ*. The measured data were corrected for self-demagnetization. The corresponding demagnetizing factors along the axial direction of the cylindrical pellets were calculated using the formula derived by Chen et al. (eq 13 in ref 46), and the corrected curves are plotted in Figure 8a.

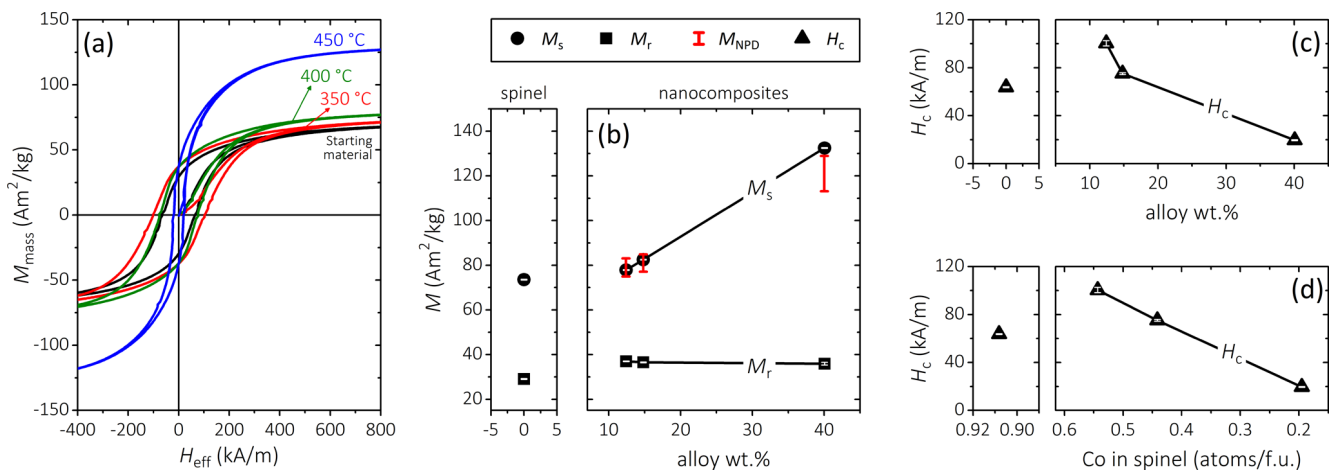

**Figure 8.** (a) Room temperature magnetic hysteresis loops measured for the starting material (black) and nanocomposites prepared *ex situ* at 350 °C (red), 400 °C (green), and 450 °C (blue). (b) Saturation magnetization, $M_s$ (black circles), remanent magnetization, $M_r$ (black squares), and calculated magnetization, $M_{NPD}$ (red bars), as a function of the reduction degree. (c) Measured coercivity, $H_c$ (black triangles), as a function of the reduction degree and (d) as a function of the amount of Co in the spinel phase in atoms per formula unit (f.u.).

**Table 2. Calculated Magnetization Value Based on the Results of Rietveld Analysis, $M_{NPD}$, Saturation Magnetization, $M_s$, Remanent Magnetization, $M_r$, Remanence-to-Saturation Ratio, $M_r/M_s$, and Coercivity, $H_c$, Extracted from the Measured Hysteresis**

| sample | $M_{NPD}$ (A m²/kg) | $M_s$ (A m²/kg) | $M_r$ (A m²/kg) | $M_r/M_s$ | $H_c$ (kA/m) |
|---|---|---|---|---|---|
| starting material | 85(4) | 73.5(2) | 29.1(2) | 0.40 | 63.7(4) |
| 350 °C | 79(4) | 77.9(2) | 37.0(2) | 0.48 | 100(2) |
| 400 °C | 81(4) | 82.5(2) | 36.5(2) | 0.44 | 75.1(7) |
| 450 °C | 121(8) | 132.5(2) | 35.9(6) | 0.27 | 19.6(4) |

The magnetic properties obtained from the hysteresis loops are displayed in Table 2 and represented in Figure 8b–d as a function of the alloy wt % and the Co content of the spinel.

The saturation magnetization, $M_s$, values were calculated using the law of approach to saturation.[47] $M_s$ increases with reduction degree following a relatively linear fashion, from 73.5(2) to 132.5(2) A m²/kg, for the starting material and the more reduced nanocomposite, respectively. This trend can be explained in terms of sample composition, as the increase in $M_s$ is directly proportional to the amount of soft phase, i.e., metallic alloy (see Figure 8b). The magnetization values extracted from Rietveld analysis of the NPD data, $M_{NPD}$, are also plotted in Figure 8b. These values were calculated as the weighted average of the atomic moments refined for the magnetic phases (spinel and alloy). See the Supporting Information for a detailed description of these calculations. The calculated $M_{NPD}$ are appreciably close in number to the measured $M_s$ values.

The remanence, $M_r$, was obtained from a linear fit of the curve near H = 0. According to Kronmüller et al. and based on the Brown–Aharoni model,[48–50] a decrease in $M_r$ is expected upon introduction of a soft material in the system, unless the soft phase is effectively exchange-coupled to the hard phase. Figure 8b reveals moderate enhancement of $M_r$ for all the nanocomposites with respect to the starting material, which indicates that the two magnetic phases must be at least partially coupled. However, an increase in $M_r$ on isotropic powders can also respond to other effects, e.g., magnetic alignment, which are not evaluated in this work. The $M_r$ of these samples are within the range expected for the $M_s$ values, according to previous studies on this system.[16,20,22,51,52]

In nanocomposites with random grain orientations, a $M_r/M_s$ above 0.5 is usually considered indicative of an effective exchange-coupling.[6] All the $M_r/M_s$ values displayed in Table 2 are below 0.5. This deviation from the theory is not entirely unexpected, as the samples prepared in this work are well apart from the ideal case (our particles are not single-domain; they present cubic anisotropy and they interact with each other). However, increased $M_r/M_s$ values are observed for the low-temperature nanocomposites with respect to the starting powders, suggesting some degree of exchange-coupling in those two samples.

The coercivity, $H_c$, was obtained from a linear fit of the curve near M = 0. Hydrothermally synthesized $CoFe_2O_4$ nanoparticles with sizes >8 nm are expected to present blocking temperatures, $T_B$, above room temperature.[27] Therefore, in the following discussion, the influence of $T_B$ on $H_c$ is neglected. In Figure 8c, $H_c$ is plotted as a function of the alloy wt %. A $H_c$ of 100(2) kA/m is observed for the nanocomposite with the lowest reduction degree (i.e., 350 °C), which implies a significant increase with respect to the starting material. The elevated temperature induces a moderate increase in crystallite size, and most likely an improvement of the crystallinity (not measured here), which boosts the $H_c$. Moreover, the monoxide (paramagnetic at room temperature, i.e., nonmagnetic) has previously been suggested to play a role in the $H_c$ of $CoFe_2O_4$/Co–Fe composites, acting as pinning sites for the domain wall.[19] The $H_c$ decreases down to 75(2) kA/m for the composite that follows in alloy wt % (i.e., 400 °C). This value is still above that of the nonreduced material, but the loss of 25 kA/m in $H_c$ seems excessive for the very small difference in alloy wt % between these two composites (alloy = 12.39(4) wt % for 350 °C and 14.80(3) wt % for 400 °C). Although the $H_c$ drop could be due, to some extent, to the crystallite size of the soft phase becoming too large to fulfill the rigid exchange-coupling condition,[53] a very clear correlation is observed between the $H_c$ and the Co content in the spinel.

As shown in Figure 8d, $H_c$ decreases linearly as the spinel Co content decreases. The $H_c$ values reported for $CoFe_2O_4$ nanoparticles are always larger than those for $\gamma$-$Fe_2O_3$ or $Fe_3O_4$ nanoparticles of comparable size and morphology.[12,54] Consequently, the loss of Co in the spinel structure inevitably causes a softening of this phase and, in turn, of the magnetic composite as a whole. Thus, the $H_c$ is dramatically diminished for the 450 °C composite as a result of the high amount of soft phase (alloy = 40.03(9) wt %) and the pronounced Co deficiency in the hard phase, $Co_{0.23(7)}Fe_{2.77(7)}O_4$. In fact, the $H_c$ measured for this sample (19.6(4) kA/m) is in the order of what is reported for pure $Fe_3O_4$ nanoparticles.[54] Comparing the $H_c$ of the composites discussed here with previous literature is not straightforward. The range of values for composites with a comparable $M_s$ is rather wide, since the effect of size and elemental composition on $H_c$ is pronounced, yet poorly analyzed in the literature.

## CONCLUSIONS

$CoFe_2O_4$ (hard)/Co—Fe alloy (soft) magnetic nanocomposites were prepared via thermal treatment of $CoFe_2O_4$ nanoparticles in the presence of $H_2$. The reduction from single-phase spinel to pure metallic alloy was followed in situ with a time resolution of 5 s using synchrotron PXRD. The in situ data revealed the appearance of a monoxide ($Co_xFe_{1-x}O$) as an intermediate phase during the reduction.

High-resolution PXRD and NPD patterns and joint Rietveld analysis of the diffraction data yielded quantitative structural and microstructural information, e.g., sample composition, crystallite size, and elemental composition of the individual phases. It was found that the reduced phases (i.e., monoxide and alloy) are rich in Co when they first emerge, at the expense of leaving a Co-deficient spinel behind. As the reduction progresses, Fe is gradually incorporated in the Co-rich phases, and toward the end of the reduction, the elemental composition of the alloy approaches the stoichiometry of the starting spinel material.

The interpretation of the refined elemental compositions in terms of the measured magnetic properties show that the Co deficiency in the spinel structure softens the magnetic material. However, our in situ investigations show that this magnetic softening may be avoided at the preparation step: lower temperatures minimize the Co deficiency in the spinel, thus diminishing the magnetic softening of the hard magnetic phase. This study provides fundamental knowledge on the reduction mechanism in $CoFe_2O_4$ systems, and helps to map parameters space, in order for tailored design of $CoFe_2O_4$/Co—Fe nanocomposite formation. The findings derived from the exhaustive characterization conducted here are also of great help in better understanding some of the observations previously reported on the topic.

## ASSOCIATED CONTENT

### Supporting Information

The Supporting Information is available free of charge on the ACS Publications website at DOI: 10.1021/acsanm.8b00808.

> Preparation of the precursor gel for hydrothermal synthesis of $CoFe_2O_4$ nanoparticles; details on Rietveld analysis of in situ and ex situ powder diffraction data; atomic scattering factors and dispersion corrections; demagnetization correction and calculation of the total sample magnetization from the refined atomic moments (PDF)

## AUTHOR INFORMATION

**Corresponding Author**
*E-mail: mch@chem.au.dk (M.C.).
**ORCID**
Cecilia Granados-Miralles: 0000-0002-3679-387X
Mogens Christensen: 0000-0001-6805-1232
**Present Address**
C.G.M.: Electroceramic Department, Instituto de Cerámica y Vidrio, CSIC, Kelsen 5, 28049 Madrid, Spain.
**Notes**
The authors declare no competing financial interest.

## ACKNOWLEDGMENTS

The authors thank financial support from the European Commission through the NANOPYME (FP7-SMALL-310516) and AMPHIBIAN (H2020-NMBP-2016-720853) projects. Financial support from the Danish National Research Foundation (Center for Materials Crystallography, DNRF-93) and the Danish Center for Synchrotron and Neutron Science (DanScatt) is gratefully acknowledged. Parts of this research were carried out at PETRA III at DESY, a member of the Helmholtz Association (HGF). We thank Jozef Bednarcik for assistance in using beamline P02.1. This work is based on experiments performed at the Swiss spallation neutron source SINQ, Paul Scherrer Institute, Villigen, Switzerland. This project has received funding from the European Union's Seventh Framework Programme for research, technological development, and demonstration under the NMI3-II Grant 283883.

## REFERENCES

(1) Balamurugan, B.; Sellmyer, D. J.; Hadjipanayis, G. C.; Skomski, R. Prospects for Nanoparticle-Based Permanent Magnets. *Scr. Mater.* **2012**, *67*, 542−547.

(2) Lewis, L. H.; Jimenez-Villacorta, F. Perspectives on Permanent Magnetic Materials for Energy Conversion and Power Generation. *Metall. Mater. Trans. A* **2013**, *44*, 2−20.

(3) Gutfleisch, O.; Willard, M. A.; Brück, E.; Chen, C. H.; Sankar, S. G.; Liu, J. P. Magnetic Materials and Devices for the 21st Century: Stronger, Lighter, and More Energy Efficient. *Adv. Mater.* **2011**, *23*, 821−842.

(4) Jimenez-Villacorta, F.; Lewis, L. H. Advanced Permanent Magnetic Materials. In *Nanomagnetism*; Estevez, J. M. G., Ed.; One Central Press: 2014; pp 161−189.

(5) Coey, J. M. D. *Magnetism and Magnetic Materials*; Cambridge University Press: New York, 2009.

(6) Kneller, E. F.; Hawig, R. The Exchange-Spring Magnet: A New Material Principle for Permanent Magnets. *IEEE Trans. Magn.* **1991**, *27*, 3588−3600.

(7) Skomski, R. Nanomagnetics. *J. Phys.: Condens. Matter* **2003**, *15*, R841−R896.

(8) Liu, J. P. Exchange-Coupled Nanocomposite Permanent Magnets. In *Nanoscale Magnetic Materials and Applications*; Liu, J. P., Fullerton, E., Gutfleisch, O., Sellmyer, D. J., Eds.; Springer: Boston, MA, 2009; pp 306−336.

(9) Zeng, H.; Li, J.; Liu, J. P.; Wang, Z. L.; Sun, S. Exchange-Coupled Nanocomposite Magnets by Nanoparticle Self-Assembly. *Nature* **2002**, *420*, 395−398.

(10) Zeng, H.; Sun, S. Syntheses, Properties, and Potential Applications of Multicomponent Magnetic Nanoparticles. *Adv. Funct. Mater.* **2008**, *18*, 391−400.

(11) Jones, N. Materials Science: The Pull of Stronger Magnets. *Nature* **2011**, *472*, 22−23.

(12) López-Ortega, A.; Lottini, E.; Fernández, C. de J.; Sangregorio, C. Exploring the Magnetic Properties of Cobalt-Ferrite Nanoparticles for the Development of a Rare-Earth-Free Permanent Magnet. *Chem. Mater.* **2015**, *27*, 4048−4056.

(13) Yue, M.; Zhang, X.; Liu, J. P. Fabrication of Bulk Nanostructured Permanent Magnets with High Energy Density: Challenges and Approaches. *Nanoscale* **2017**, *9*, 3674−3697.

(14) Li, X.; Lou, L.; Song, W.; Zhang, Q.; Huang, G.; Hua, Y.; Zhang, H.-T.; Xiao, J.; Wen, B.; Zhang, X. Controllably Manipulating Three-Dimensional Hybrid Nanostructures for Bulk Nanocomposites with Large Energy Products. *Nano Lett.* **2017**, *17*, 2985−2993.

(15) de Assis Olimpio Cabral, F.; de Araujo Machado, F. L.; de Araujo, J. H.; Soares, J. M.; Rodrigues, A. R.; Araujo, A. Preparation and Magnetic Study of the $CoFe_2O_4$-$CoFe_2$ Nanocomposite Powders. *IEEE Trans. Magn.* **2008**, *44*, 4235−4238.

(16) Soares, J. M.; Cabral, F. A. O.; de Araújo, J. H.; Machado, F. L. A. Exchange-Spring Behavior in Nanopowders of $CoFe_2O_4$−$CoFe_2$. *Appl. Phys. Lett.* **2011**, *98*, 072502.

(17) Quesada, A.; Rubio-Marcos, F.; Marco, J. F.; Mompean, F. J.; García-Hernández, M.; Fernández, J. F. On the Origin of Remanence Enhancement in Exchange-Uncoupled $CoFe_2O_4$-Based Composites. *Appl. Phys. Lett.* **2014**, *105*, 202405.

(18) Jin, J.; Sun, X.; Wang, M.; Ding, Z. L.; Ma, Y. Q. The Magnetization Reversal in $CoFe_2O_4$/$CoFe_2$ Granular Systems. *J. Nanopart. Res.* **2016**, *18*, 383.

(19) Quesada, A.; Granados-Miralles, C.; López-Ortega, A.; Erokhin, S.; Lottini, E.; Pedrosa, J.; Bollero, A.; Aragón, A. M.; Rubio-Marcos, F.; Stingaciu, M.; Bertoni, G.; de Julián Fernández, C.; Sangregorio, C.; Fernández, J. F.; Berkov, D.; Christensen, M. Energy Product Enhancement in Imperfectly Exchange-Coupled Nanocomposite Magnets. *Adv. Electron. Mater.* **2016**, *2*, 1500365.

(20) Leite, G. C. P.; Chagas, E. F.; Pereira, R.; Prado, R. J.; Terezo, A. J.; Alzamora, M.; Baggio-Saitovich, E. Exchange Coupling Behavior in Bimagnetic $CoFe_2O_4$/$CoFe_2$ Nanocomposite. *J. Magn. Magn. Mater.* **2012**, *324*, 2711−2716.

(21) Li, W.; Liu, Y.; Guo, Y. The Synthesis of Magnetically Exchange Coupled CoFe2O4/CoFe2 Composites through Low Temperature CaH2 Reduction. *Mater. Res. Express* **2017**, *4*, 085019.

(22) Zan, F. L.; Ma, Y. Q.; Ma, Q.; Zheng, G. H.; Dai, Z. X.; Wu, M. Z.; Li, G.; Sun, Z. Q.; Chen, X. S. One-Step Hydrothermal Synthesis and Characterization of High Magnetization $CoFe_2O_4$/$Co_{0.7}Fe_{0.3}$ Nanocomposite Permanent Magnets. *J. Alloys Compd.* **2013**, *553*, 79−85.

(23) Zhang, Y.; Xiong, R.; Yang, Z.; Yu, W.; Zhu, B.; Chen, S.; Yang, X. Enhancement of Interparticle Exchange Coupling in $CoFe_2O_4$/$CoFe_2$ Composite Nanoceramics Via Spark Plasma Sintering Technology. *J. Am. Ceram. Soc.* **2013**, *96*, 3798−3804.

(24) Ou-Yang, J.; Zhang, Y.; Luo, X.; Yan, B.; Zhu, B.; Yang, X.; Chen, S. Composition Dependence of the Magnetic Properties of $CoFe_2O_4$/$CoFe_2$ Composite Nano-Ceramics. *Ceram. Int.* **2015**, *41*, 3896−3900.

(25) Soares, J. M.; Conceição, O. L. A.; Machado, F. L. A.; Prakash, A.; Radha, S.; Nigam, A. K. Magnetic Couplings in $CoFe_2O_4$/FeCo−FeO Core−Shell Nanoparticles. *J. Magn. Magn. Mater.* **2015**, *374*, 192.

(26) Sears, V. F. Scattering Lengths for Neutrons. In *International Tables for Crystallography Vol C: Mathematical, Physical and Chemical Tables*; Prince, E., Ed.; Kluwer Academic Publishers: Dordrecht, The Netherlands, 2004; pp 444−454.

(27) Stingaciu, M.; Andersen, H. L.; Granados-Miralles, C.; Mamakhel, A.; Christensen, M. Magnetism in CoFe2O4 Nanoparticles Produced at Sub- and near-Supercritical Conditions of Water. *CrystEngComm* **2017**, *19*, 3986−3996.

(28) Andersen, H. L.; Christensen, M. In Situ Powder X-Ray Diffraction Study of Magnetic $CoFe_2O_4$ Nanocrystallite Synthesis. *Nanoscale* **2015**, *7*, 3481−3490.

(29) Dippel, A. C.; Liermann, H. P.; Delitz, J. T.; Walter, P.; Schulte-Schrepping, H.; Seeck, O. H.; Franz, H. Beamline P02.1 at PETRA III for High-Resolution and High-Energy Powder Diffraction. *J. Synchrotron Radiat.* **2015**, *22*, 675−687.

(30) Prescher, C.; Prakapenka, V. B. DIOPTAS: A Program for Reduction of Two-Dimensional X-Ray Diffraction Data and Data Exploration. *High Pressure Res.* **2015**, *35*, 223.

(31) NIST. Standard Reference Material 660b: Line Position and Line Shape Standard for Powder Diffraction; https://www-s.nist.gov/ (accessed Feb 6, 2017).

(32) Andersen, H. L.; Bøjesen, E. D.; Birgisson, S.; Christensen, M.; Iversen, B. B. Pitfalls and Reproducibility of in Situ Synchrotron Powder X-Ray Diffraction Studies of Solvothermal Nanoparticle Formation. *J. Appl. Crystallogr.* **2018**, *51*, 526−540.

(33) Schefer, J.; Fischer, P.; Heer, H.; Isacson, A.; Koch, M.; Thut, R. A Versatile Double-Axis Multicounter Neutron Powder Diffractometer. *Nucl. Instrum. Methods Phys. Res., Sect. A* **1990**, *288*, 477−485.

(34) Fischer, P.; Frey, G.; Koch, M.; Könnecke, M.; Pomjakushin, V.; Schefer, J.; Thut, R.; Schlumpf, N.; Bürge, R.; Greuter, U.; Bondt, S.; Berruyer, E. High-Resolution Powder Diffractometer HRPT for Thermal Neutrons at SINQ. *Phys. B* **2000**, *276−278*, 146−147.

(35) Rodríguez-Carvajal, J. Recent Advances in Magnetic Structure Determination by Neutron Powder Diffraction. *Phys. B* **1993**, *192*, 55−69.

(36) Granados-Miralles, C.; Saura-Múzquiz, M.; Bøjesen, E. D.; Jensen, K. M. Ø.; Andersen, H. L.; Christensen, M. Unraveling Structural and Magnetic Information during Growth of Nanocrystalline $SrFe_{12}O_{19}$. *J. Mater. Chem. C* **2016**, *4*, 10903−10913.

(37) Ohnuma, I.; Enoki, H.; Ikeda, O.; Kainuma, R.; Ohtani, H.; Sundman, B.; Ishida, K. Phase Equilibria in the Fe−Co Binary System. *Acta Mater.* **2002**, *50*, 379−393.

(38) Willis, B. T. M.; Rooksby, H. P. Change of Structure of Ferrous Oxide at Low Temperature. *Acta Crystallogr.* **1953**, *6*, 827−831.

(39) Yamamoto, A. Modulated Structure of Wustite ($Fe_{1−x}$O) (Three-Dimensional Modulation). *Acta Crystallogr., Sect. B: Struct. Crystallogr. Cryst. Chem.* **1982**, *38*, 1451−1456.

(40) Fjellvåg, H.; Grønvold, F.; Stølen, S.; Hauback, B. On the Crystallographic and Magnetic Structures of Nearly Stoichiometric Iron Monoxide. *J. Solid State Chem.* **1996**, *124*, 52−57.

(41) Redman, M. J.; Steward, E. G. Cobaltous Oxide with the Zinc Blende/Wurtzite-Type Crystal Structure. *Nature* **1962**, *193*, 867.

(42) Tombs, N. C.; Rooksby, H. P. Structure of Monoxides of Some Transition Elements at Low Temperatures. *Nature* **1950**, *165*, 442−443.

(43) Liu, J. F.; Yin, S.; Wu, H. P.; Zeng, Y. W.; Hu, X. R.; Wang, Y. W.; Lv, G. L.; Jiang, J. Z. Wurtzite-to-Rocksalt Structural Transformation in Nanocrystalline CoO. *J. Phys. Chem. B* **2006**, *110*, 21588−21592.

(44) Vanýsek, P. Electrochemical Series. In *CRC Handbook of Chemistry and Physics*, Internet Version 2006; Lide, D. R., Ed.; Taylor and Francis: Boca Raton, FL, 2006.

(45) Jensen, K. M. Ø.; Andersen, H. L.; Tyrsted, C.; Bøjesen, E. D.; Dippel, A.-C.; Lock, N.; Billinge, S. J. L.; Iversen, B. B.; Christensen, M. Mechanisms for Iron Oxide Formation under Hydrothermal Conditions: An in Situ Total Scattering Study. *ACS Nano* **2014**, *8*, 10704−10714.

(46) Chen, D. X.; Pardo, E.; Sanchez, A. Fluxmetric and Magnetometric Demagnetizing Factors for Cylinders. *J. Magn. Magn. Mater.* **2006**, *306*, 135−146.

(47) Zhang, H.; Zeng, D.; Liu, Z. The Law of Approach to Saturation in Ferromagnets Originating from the Magnetocrystalline Anisotropy. *J. Magn. Magn. Mater.* **2010**, *322*, 2375−2380.

(48) Brown, W. F. Virtues and Weaknesses of the Domain Concept. *Rev. Mod. Phys.* **1945**, *17*, 15−19.

(49) Aharoni, A. Theoretical Search for Domain Nucleation. *Rev. Mod. Phys.* **1962**, *34*, 227−238.

(50) Kronmüller, H. Theory of Nucleation Fields in Inhomogeneous Ferromagnets. *Phys. Phys. Status Solidi B* **1987**, *144*, 385−396.

(51) Soares, J. M.; Galdino, V. B.; Conceição, O. L. A.; Morales, M. A.; de Araújo, J. H.; Machado, F. L. A. Critical Dimension for Magnetic Exchange-Spring Coupled Core/Shell $CoFe_2O_4$/$CoFe_2$ Nanoparticles. *J. Magn. Magn. Mater.* **2013**, *326*, 81−84.

(52) Soares, J. M.; Galdino, V. B.; Machado, F. L. A. Exchange-Bias and Exchange-Spring Coupling in Magnetic Core−shell Nanoparticles. *J. Magn. Magn. Mater.* **2014**, *350*, 69−72.

(53) Fullerton, E. E.; Jiang, J.; Bader, S. Hard/Soft Magnetic Heterostructures: Model Exchange-Spring Magnets. *J. Magn. Magn. Mater.* **1999**, *200*, 392−404.

(54) Li, Q.; Kartikowati, C. W.; Horie, S.; Ogi, T.; Iwaki, T.; Okuyama, K. Correlation between Particle Size/Domain Structure and Magnetic Properties of Highly Crystalline Fe$_3$O$_4$ Nanoparticles. *Sci. Rep.* **2017**, *7*, 9894.